\documentclass[runningheads]{svmult}
\usepackage{epsfig}
\usepackage{makeidx}   
\usepackage{graphicx}  
\usepackage{subeqnar}  
\usepackage{multicol}  
\usepackage{physprbb}  
\makeindex             

\begin{document}
\title*{Quantum states of neutrons in the gravitational field and limits for non-Newtonian interaction in the range between 1 $\micro$m and 10 $\micro$m}
\toctitle{Quantum states\protect\newline}
%
%
\titlerunning{Quantum states in the gravitational field}
%
\author{Hartmut Abele\inst{1}, Stefan Bae$\ss{}$ler\inst{2} and Alexander Westphal\inst{3}}
\authorrunning{H. Abele, S. Bae$\ss$ler and A. Westphal}
%
%
\institute{Physikalisches Institut der Universit\"at Heidelberg,\\
Philosophenweg 12, 69120 Heidelberg, Germany \and Physikalisches Institut der Universit\"at Mainz,\\
Staudingerweg 7, 55099 Mainz, Germany  \and DESY Hamburg,
\\Notkestra$\ss$e 85, 22607 Hamburg, Germany}

\maketitle              

\begin{abstract}
Quantum states in the Earth's gravitational field can be observed,
when ultra-cold neutrons fall under gravity. In an experiment at
the Institut Laue-Langevin in Grenoble, neutrons are reflected and
trapped in a gravitational cavity above a horizontal mirror. The
population of the ground state and the lowest states follows, step
by step, the quantum mechanical prediction. An efficient neutron
absorber removes the higher, unwanted states. The quantum states
probe Newtonian gravity on the micrometer scale and we place
limits for gravity-like forces in the range between 1 $\micro$m
and 10 $\micro$m.
\end{abstract}

\section{A quantum system}
Quantum physics is a fascinating but subtle subject. The subtlety
of the quantum system described here arises from the fact, that
gravity appears very weak in our universe. Quantum theory and
gravitation affect each other, and, when neutrons become
ultra-cold, the fall experiment of Galileo Galilei shows quantum
aspects of the subtle gravity force in that sense that neutrons do
not fall continuously. We find them on different levels, when they
come close to a reflecting mirror for
neutrons~\cite{Nesvizhevsky1}. Of course, such bound states with
discrete energy levels are expected when the graviational
potential is larger than the energy of the particle. Here, the
quantum states have pico-eV energy, a value that is smaller by
many orders of magnitude compared with an electromagnetically
bound electron in a hydrogen atom, opening the way to a new
technique for gravity experiments, for neutron optics, neutron
detection and measurements of fundamental properties.

A side-effect of this experiment is its sensitivity for
gravity-like forces at length scales below 10 $\micro$m. In light
of recent theoretical developments in higher dimensional field
theory~\cite{Arkani1,Arkani2,Antoniadis}, gauge fields can mediate
forces that are $10^{10}$ to $10^{12}$ times stronger than gravity
at submillimeter distances, exactly in the interesting range of
this experiment and might give a signal in an improved setup.

The idea of observing quantum effects occuring when ultracold
neutron are stored on a plane was discussed long ago by V.I.
Lushikov and I.A. Frank~\cite{Lushikov}. An in some aspects
similar experiment was discussed by H. Wallis et al.~\cite{Wallis}
in the context of trapping atoms in a gravitational cavity.
Retroreflectors for atoms have used the electric dipole force in
an evanescent light wave~\cite{Aminoff,Kasevich} or they are based
on the gradient of the magnetic dipole interaction, which has the
advantage of not requiring a laser~\cite{Roach}. A neutron mirror
makes use of the strong interaction between nuclei and an
ultracold neutron, resulting in an effective repulsive force:
Neutrons propagate in condensed matter in a matter similar to the
propagation of light but with a neutron refractive index less than
unity. Thus, one considers the surface of matter as constituting a
potential step of height $V$. Neutrons with transversal energy
$E_{\perp}$ $<$ $V$ will be totally reflected. Ultra-cold neutrons
(UCN) are neutrons that, in contrast to faster neutrons, are
retro-reflected from surfaces at all angles of incidence. When the
surface roughness of the mirror is small enough, the UCN
reflection is specular.

Neutron mirrors are interesting because they can be used to store
neutrons, to focus neutrons, or even to build a Fabry Perot
interferometer for neutron de Broglie waves. UCN storage bottle
experiments have improved our knowledge about the neutron lifetime
significantly or, together with the Ramsey method of separated
oscillating fields, they have been used for a search for an
electric dipole moment of the neutron.
\section{Limits for
non-Newtonian interaction below 10 $\micro$m} Discussions about
deviations from the inverse square law for gravity have become
popular in the past few years~\cite{Fischbach}. A new effective
interaction coexisting with gravity would modify the Newtonian
potential. On the assumption that the form of the non-Newtonian
potential is given by the Yukawa expression, for masses $m_i$ and
$m_j$ and distance $r$ the modified Newtonian potential $V(r)$ is
having the form
\begin{equation}
V(r) = -G\frac{m_i\cdot{m_j}}{r}(1+\alpha\cdot{e^{-r/\lambda}})
\end{equation}
\begin{figure}
\begin{center}
\includegraphics{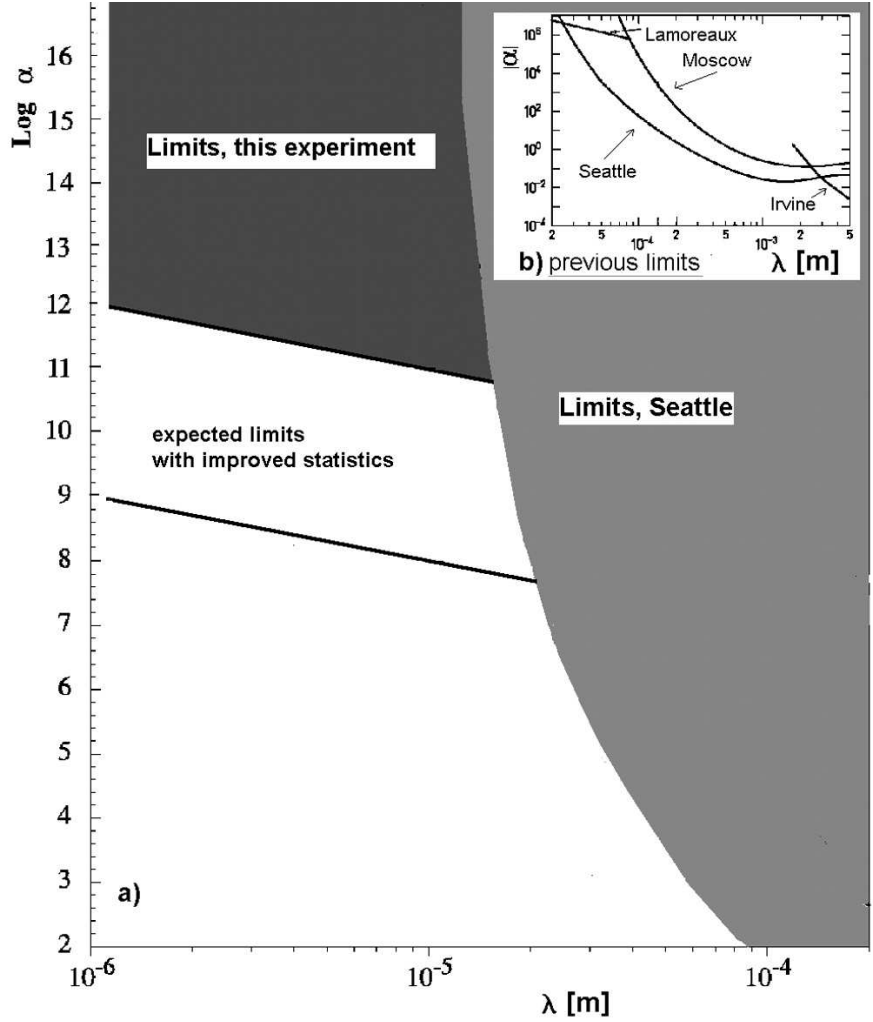}
\end{center}
\caption[]{Limits for non-Newtonian gravity: Strength $|\alpha|$
vs. Yukawa length scale $\lambda$\\
a) Experiments with neutrons place limits for $|\alpha|$ in the
range 1 $\micro$m $<$ $\lambda$ $<$ 10 $\micro$m.\\ b) Constraints
from previous
experiments~\cite{Lamoreaux,Hoskins,Mitrofanov,Hoyle,Gundlach} are
adapted from Ref.~\cite{Hoyle}.} \label{eps1}
\end{figure}where $\lambda$ is the Yukawa distance over which the
corresponding force acts and $\alpha$ is a strength factor in
comparison with Newtonian gravity. $G$ is the gravitational
constant. For a neutron with mass $m_n$, the gravitational force
of the mass $m_E$ of the entire earth with radius $R_E$ lead to a
free fall acceleration
\begin{equation}
g = \frac{Gm_E}{R_E^2}=\frac{4\pi\cdot{G\rho
R_E^3}}{3R_E^2}=\frac{4}{3}\pi{G\rho R_E}.
\end{equation}
When a neutron approaches the mirror, the mass of this extended
source might modify $g$, when strong non-Newtonian forces are
present. For small neutron distances $z$ from the mirror, say
several micrometers, we consider the mirror as an infinite
half-space with mass density $\rho$. By replacing the source mass
$m_i$ by $dm_i$ and integrating over $dm_i$, the  Yukawa-like term
$\lambda$ modified Newtonian potential $V^{\prime}(r)$ is having
the form
\begin{equation}
V^{\prime}(z,\lambda) = 2\pi\cdot{m_n\rho}\alpha\lambda^2
G\cdot{e^{-|z|/\lambda}}. \end{equation} and the non-Newtonian
acceleration $g^{\prime}$ as a function of distance $z$ is given
by
\begin{equation}
g^{\prime}(z,\lambda) = 2\pi\cdot{\rho}\alpha\lambda
G\cdot{e^{-|z|/\lambda}}
\end{equation}
As a consequence, the ratio is
\begin{equation}
\frac{g^{\prime}}{g}(z,\lambda) =
\frac{3}{2}\alpha\cdot{\frac{\lambda}{R_E}}\cdot{e^{-|z|/\lambda}}.
\end{equation}For z = $\lambda$ = 10 $\mu$m and $\alpha$ = 1, the ratio
$g^{\prime}$/$g$ is about 10$^{-12}$. Fig. 1a shows the present
status of an experimental search for gravity-like forces below 10
$\micro$m. The results of a fit of potential Eq. 3 to the measured
data (see Fig. 5 and Fig. 6) yields predictions for 90\%
confidence level exclusion bounds on $\alpha$ and $\lambda$. These
limits from this neutron mirror experiment are shown in Fig.1a.
They are the best known in the range 1 $\micro$m $<$ $\lambda$ $<$
3 $\micro$m and exclude for the first time gravity-like
short-ranged forces at 1 $\micro$m with strength $\alpha$ $>$
10$^{12}$. The limit for strength $\alpha$ at 10 $\micro$m is
10$^{11}$ (Fig. 1a)). There is a theoretical uncertainty in our
limits since they depend on the model for the neutron absorber,
estimated to be about one order of magnitude. In future
experiments, these limits will be strongly improved by an enhanced
setup and improved statistics by new UCN sources as a Monte Carlo
simulation shows. Previous
constraints~\cite{Lamoreaux,Hoskins,Mitrofanov,Hoyle,Gundlach} on
both $\lambda$ and $\alpha$ are shown in Fig. 1b.

The method with neutrons has some advantages. Electromagnetic
interactions at micrometer distances, serious sources of
systematic error in distance force measurements, are effectively
suppressed. The neutron carries no electric charge and direct
electrostatic forces are ruled out. Yet, it bears a very tiny
magnetic moment $\mu_n$ of roughly $0.5\cdot 10^{-3}\cdot\mu_B$.
This magnetic moment can induce a magnetic mirror force onto a
neutron that hovers close to the surface of a body. Further, a
neutron moving with the velocity $v$ sees an induced electrostatic
force, that is essentially some kind of Lorentz force and thus an
effect of the order of $v/c$. Both effects can be evaluated to
yield electrodynamic energy shifts. With permittivity $\epsilon$
and permeability $\mu$, the order of magnitude is
\[\Delta E_{\vec{E},v}\sim\epsilon_0\cdot\frac{\epsilon-1}{\epsilon}\cdot\frac{\mu_0^2\cdot\mu_n^2}{48\pi}\cdot\frac{v^2}{z^3}\sim10^{-26}\cdot10^{-12}\,eV\]
\[\Delta E_{\vec{B}}\sim\frac{\mu-1}{\mu}\cdot\frac{\mu_0}{16\pi}\cdot\frac{\mu_n^2}{z^3} \sim10^{-13}\cdot10^{-12}\,eV\]for $v\sim 10\,m/s$ and $z\sim 10\,\mu m$. Thus, these effects can completely be neglected, since they are by far subgravitational in strength.

Motivations for gravity experiments come from frameworks for
solving the hierarchy problem in a way of bringing quantum gravity
down to the TeV scale. In such frameworks the Standard Model
fields are localized on a 3-brane in a higher dimensional space by
the presence of new dimensions of submillimeter
size~\cite{Arkani1}. At the expected sensitivity, a number of
gravity-like phenomena emerge. For example, a hypothetical gauge
field can naturally have miniscule gauge coupling $g_4$ $\sim$
10$^{-16}$ for 1 TeV, independent of the number of extra
dimensions~\cite{Arkani2}. If these gauge fields couple to a
neutron with mass $m_n$, the ratio of the repulsive force mediated
by this gauge field to the gravitational attraction
is~\cite{Arkani2}
\begin{equation}
\frac{F_{gauge}}{F_{grav}}\sim\frac{g_4^2}{Gm_n^2}\sim10^6(\frac{g_4}{10^{-16}})^2.
\end{equation}

With $g_4$ = 10$^{-16}$ as a lower bound, these gauge fields can
result in repulsive forces of million or billion times stronger
than gravity at micrometer distances, exactly in the range of
interest.

\section{The experiment at the Institut Laue-Langevin}
\subsection{From hot to ultracold}
Neutrons are produced in a spallation source or a research
reactor. At production, these neutrons are very hot; the energy is
about 2 MeV corresponding to 10$^{10}$ degrees Centigrade. On the
other side of the scale, the gravity experiment uses neutrons
having 10$^{18}$ times less energy in the pico-eV range (see Tab.
1). In a first step, spallation or fission neutrons thermalize in
a heavy water tank at a temperature of 300 K. The thermal fluxes
are distributed in energy according to Maxwellian law. At the
Institut Laue-Langevin (ILL), cold neutrons are obtained in a
second moderator stage from a 25 K liquid deuterium cold moderator
near the core of the 57 MW uranium reactor. These cold neutrons
have a velocity spectrum in the milli-eV energy range. For
particle physics, a new beam line with a flux of more than
10$^{10}$ cm$^{-2}$s$^{-1}$ over a cross section of 6 cm x 20 cm
is available.

Ultra-cold neutrons are taken from the low energy tail of the cold
Maxwellian spectrum. They are guided vertically upwards by a
neutron guide (Fig. 2). The curved guide, which absorbs neutrons
above a threshold energy,  acts as a low-velocity filter for
neutrons. Neutrons with a velocity of up to 50 m/s arrive at a
rotating nickel turbine. Colliding with the moving blades of the
turbine, ultra-cold neutrons exit the turbine with a velocity of
several meters per second. They are then guided to several
experimental areas.
\begin{figure}
\begin{center}
\includegraphics{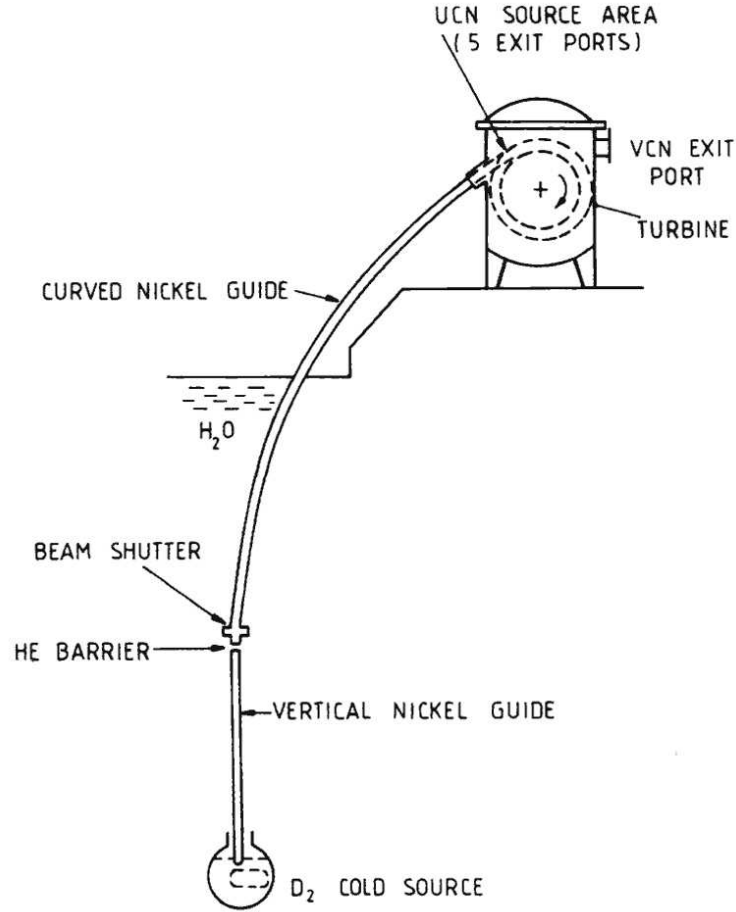}
\end{center}
\caption[]{ucn source} \label{eps2}
\end{figure}
The exit window of the guide for the gravity experiment has a
rectangular shape with the dimensions of 100 mm x 10 mm. At the
entrance of the experiment, a collimator absorber system cuts down
on the neutrons to a adjustable transversal velocity corresponding
to an energy in the pico-eV range.
\begin{table}
\caption{from hot to ultracold: neutrons at the ILL}
\begin{center}
\renewcommand{\arraystretch}{1.4}
\setlength\tabcolsep{5pt}
\begin{tabular}{lllllll}
\hline\noalign{\smallskip} & ${\mathrm {fission}}$ & ${\mathrm
{thermal}}$ & ${\mathrm {cold}}$ & ${\mathrm
{ultracold}}$&${\mathrm {this}}$\\
& ${\mathrm {neutrons}}$ &
${\mathrm {neutrons}}$ & ${\mathrm {neutrons}}$ & ${\mathrm {neutrons}}$&${\mathrm {experiment}}$\\
 \hline\noalign{\smallskip}
Energy & 2 MeV & 25 meV & 3 meV & 100 neV & 1.4 peV \\
Temperature & 10$^{10}$ K & 300 K & 40 K & 1 mK & - \\
Velocity & 10$^7$ m/s & 2200 m/s & 800 m/s & 5 m/s & $v_{\perp}\sim$ 2 cm/s\\
\hline
\end{tabular}
\end{center}
\label{apptab1b}
\end{table}
\subsection{The setup}
Fig. 3 shows a schematic view of the setup: Neutrons pass through
the mirror absorber system eventually detected by a
$^3$He-counter. The experiment itself is mounted on a polished
plane granite stone with a passive antivibration table underneath.
This stone is leveled with piezo translators. Inclinometers
together with the piezo translators in a closed loop circuit
guarantee leveling with an absolute precision better than 10
$\micro$rad. Either one solid block with dimensions 10 cm x 10 cm
x 3 cm or two solid blocks with dimensions 10 cm x 6 cm x 3 cm
composed of optical glass serve as mirrors for UCN neutron
reflection.  Small angle X-ray studies~\cite{Westphal} determined
the roughness of the surface to be $\sigma$ = 2.2 $\pm$ 0.2 nm and
the associated lateral correlation length to be $\zeta$ = 10 $\pm$
2 $\micro$m.
\begin{figure}
\begin{center}
\includegraphics{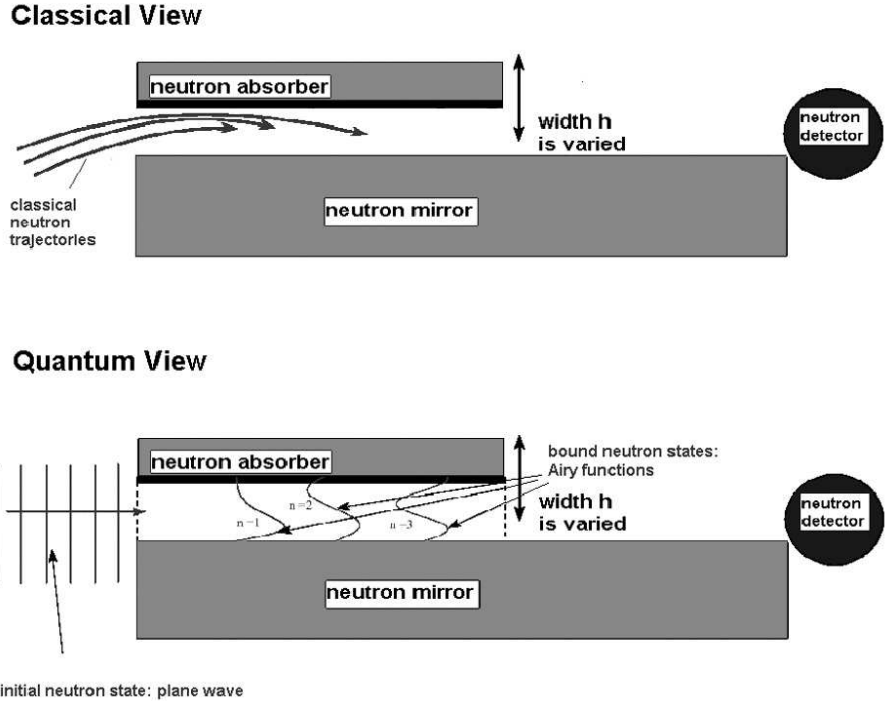}
\end{center}
\caption[]{Sketch of the setup: a) classical view: neutron
trajectories, b) quantum view: plane waves and Airy functions}
\label{eps1}
\end{figure}
The plane can thus be regarded as a pattern that varies in height
with 2 nm on a scale of 10 $\micro$m. Since the de Broglie
wavelength of the neutrons is in the range of 40nm to 100nm, the
neutrons do see a surface that is essentially flat. A
neutron-absorber is placed above the first mirror. The absorber
consists of a rough glass plate coated with an Gd-Ti-Zr alloy by
means of magnetron evaporation. The absorbing layer is 200nm
thick. The surface of the absorber was parallel to the surface of
the mirror. The absorber roughness and correlation length was
measured with an atomic force microscope to be $\sigma$ = 0.75
$\micro$m and $\zeta$ = 5 $\micro$m respectively. Neutrons that
hit the absorber surface are either absorbed in the alloy or
scattered out of the experiment at large angles. The efficiency of
removing these fast unwanted neutrons is 93\%. The collimation
system in front of the mirror absorber system is adjusted in that
way, that classical trajectories of neutrons entering the
experiment have to hit the mirror surface at least two times.
After the second mirror we placed a $^{3}$He counter for neutron
detection. More information about the setup can be found
in~\cite{Nesvizhevsky2}.

\section{Gravity and quantum mechanics work together}
\subsection{Theoretical description} The
neutrons fall under gravity onto the mirror. The calculation of
the energy eigenvalues of the vertical motion of the neutrons in
the mirror-absorber system is a nice example of quantum mechanics.
In fact, we have two theoretical descriptions for the transmission
of neutrons. The first one is the well known WKB method. Usually,
the accuracy of WKB quantization is 20\% for the ground state,
whereas the accuracy increases for higher levels. A similar
calculation of energy levels for the gravitational field with the
WKB method can also be found in~\cite{Wallis}. We can compare the
WKB result with an exact analytical solution using Airy-functions.
Taking the neutron-absorber into account, the agreement of the two
methods is significantly better than 10\%.
\begin{table}
\caption{Eigenenergies and classical turning points for neutrons,
atoms and electrons, a comparison}
\begin{center}
\renewcommand{\arraystretch}{1.4}
\setlength\tabcolsep{5pt}
\begin{tabular}{llllll}
\hline\noalign{\smallskip} & ${\mathrm {Neutron}}$ &
$^{4}{\mathrm {Helium}}$ & $^{85}{\mathrm {Rubidium}}$ & $^{133}{\mathrm {Cesium}}$&${\mathrm {Electron}}$\\
\noalign{\smallskip} \hline \noalign{\smallskip}
$E_1\; [peV]$ & 1.4 & 38.4 & 35.7 & 154 & 0.11 \\
$E_2\; [peV]$ & 2.5 & 42.1 & 34.7 & 138 & 0.20 \\
$z_1\; [\mu]$ & 13.72 & 5.5 & 0.7 & 0.5& 2061\\
$z_2\; [\mu]$ & 23.99& 9.5 & 1.2 & 0.9 & 3604\\

\hline
\end{tabular}
\end{center}
\end{table}
We start calculations from the one dimensional stationary
Schr\"odinger equation,
\begin{equation}
-\frac{\hbar^{2}}{2m}\bigtriangleup\Psi+V(z)=E\Psi
\end{equation}
with wave function $\Psi$ for energy $E$ and the potential
\begin{subeqnarray}
V(z)& = mgz\;  {\mathrm{for}}\; z \geq 0\; , \nonumber \\
V(z)& = \infty\; {\mathrm {for}}\; z < 0, \setcounter{eqsubcnt}{0}
\end{subeqnarray}
ignoring the absorber for now. $m$ is the mass of the neutron and
$g$ is the acceleration in the earth's gravitational field. The
quantum mechanical treatment of reflecting neutron mirror, made
from glass, is simple. The glass potential is essential real
because of the small absorption cross section of glass and with
$V$ = 100 $neV$ large compared with transversal energy
$E_{\perp}$. Therefore, the potential $V$ is set to infinity at z
= 0. The quantum mechanical description follows in
part~\cite{Wallis}. It is convenient to use a scaling factor
\begin{equation}
\zeta=\frac{z}{z_0}\; {\mathrm {with}}\;
z_0={(\frac{\hbar^2}{2m^2g})}^{1/3}.
\end{equation}
Solutions of Equ. 1 for $\Psi$  are obtained with an Airy function
\begin{equation}
\Psi_{n}(\zeta)=Ai(\zeta-\zeta_{n})
\end{equation}
The displacement $\zeta_n$ of the n-th eigenvector has to coincide
with the n-th zero of the Airy function (Ai(-$\zeta_n$)=0) to
fulfill the boundary condition $\Psi_n(0)$=0 at the mirror.
Eigenfunctions (n$>$0) are
\begin{equation}
Ai(\zeta-\zeta_n)
 \setcounter{eqsubcnt}{0}
 \end{equation}
with corresponding eigenenergies
\begin{equation}
E_n=mgz_n \setcounter{eqsubcnt}{0}
\end{equation} and
\begin{equation}
z_n=z_0((\frac{3\pi}{2}(n-\frac{1}{4}))^2)^{1/3}
\setcounter{eqsubcnt}{0}
\end{equation} $z_n$ corresponds to the
turning point of a classical neutron trajectory with energy $E_n$.
For example,  Energies of the lowest levels (n = 1, 2, 3, 4) are
1.44 peV, 2.53 peV 3.42 peV and 4.21 peV. The corresponding
classical turning points $z_n$ are 13.7 $\micro$m, 24.1 $\micro$m,
32.5 $\micro$m and 40.1 $\micro$m (see Tab. 2).

\begin{equation}
\rho = C\Psi^{*}\Psi \end{equation} is the neutron density and can
be interpreted as the probability to detect a neutron at height z
above the mirror, see Fig. 4. C is a constant.

\begin{figure}
\begin{center}
\includegraphics{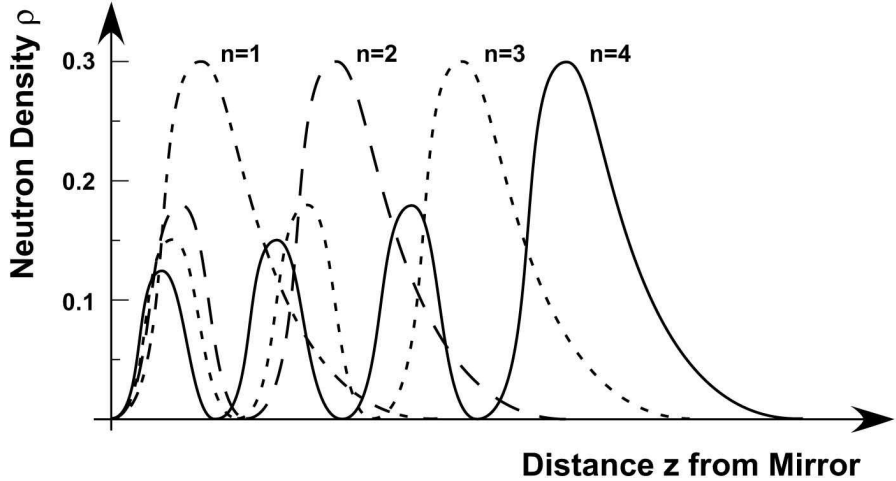}
\end{center}
\caption[]{Neutron density above the mirror for states \#1 to \#4}
\label{eps2}
\end{figure}
In principle it is possible to visualize this neutron density
distribution. The distribution is measurable with a nuclear track
neutron detector having at the moment a spatial resolution of
about 3 to 4 $\micro$m~\cite{Ruess}. The nuclear track detectors
is made out of CR39 plastic coated with 5mg/cm$^2$ $^{235}$UF$_4$.
Nuclear fission converts a neutron into a detectable track on
CR39. The tracks can be visualized with a standard optical
microscope after chemical treatment. The typical diameter of such
a track is around 1.5 $\micro$m with a length of about 10
$\micro$m. Competing reactions from $\gamma$ rays or alpha
particles have a smaller track signature and as a consequence,
background from these reactions is practically zero. The automatic
readout of the CR39 detector was done in the CHORUS group at
C.E.R.N.~\cite{Ruess}. The microscope MICOS2 is normally used to
scan radiated emulsion plates in a search for neutrino
oscillation. The rectangular stage of the microscope can be moved
by step motors with a reproducibility of 1 $\micro$m. The focal
length of the microscope is adapted to a CCD camera. The
resolution in terms of one pixel is approximately 0.34 $\micro$m.
An image analysis program detects the tracks on CR39. Having
followed the tracks in depth of CR39, the impact point of the
fission product on CR39 was found and the spatial resolution of
the detector was significantly improved.

The population of the ground state and lowest state follows the
quantum mechanical prediction. Higher, unwanted states are removed
by the rough neutron absorber made up of an alloy of Ti, Zn and
Gd.
\begin{figure}
\begin{center}
\includegraphics{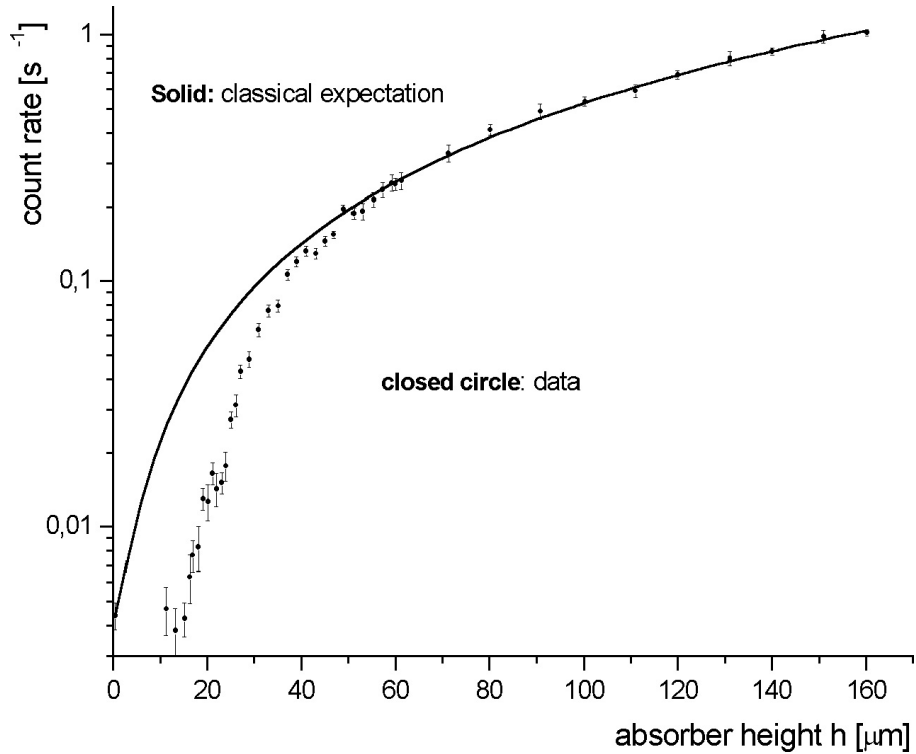}
\end{center}
\caption[]{Data vs. classical expectation} \label{eps2}
\end{figure}

\subsection{Observation of quantum states}

Signatures of quantum states in the gravitational field of the
earth are observed in the following way: A $^{3}$He counter
measures the total neutron transmission $T$, when neutrons are
traversing the mirror absorber-system as described in section 2.
The transmission is measured as a function of the absorber height
$h$ and thus as a function of neutron energy since the height acts
as a selector for the vertical energy component $E_{\perp}$, see
Fig. 3). The solid data points, plotted in Fig. 5, show the
measured number of transmitted neutrons for an absorber height $h$
from zero up to 160 $\micro$m. From the classical point of view,
the transmission $T$ of neutrons is proportional to the phase
space volume allowed by the absorber. It is governed by a power
law $T$ $\sim$ $h^n$ and n = 1.5 The solid line in Fig. 5 shows
this classical expectation.

Above an absorber height of about 60$\micro$m, the
measured transmission is in agreement with the classical
expectation but below 50 $\micro$m, a deviation is clearly
visible. From quantum mechanics, we easily understand this
behavior: Ideally, we expect a stepwise dependence of $T$ as a
function of $h$. If $h$ is smaller than the spatial width of the
lowest quantum state, then $T$ will be zero. When $h$ is equal to
the spatial width of the lowest quantum state then $T$ will
increase sharply. A further increase in $h$ should not increase
$T$ as long as $h$ is smaller than the spatial width of the second
quantum state. Then again, $T$ should increase stepwise.
\begin{figure}
\begin{center}
\includegraphics{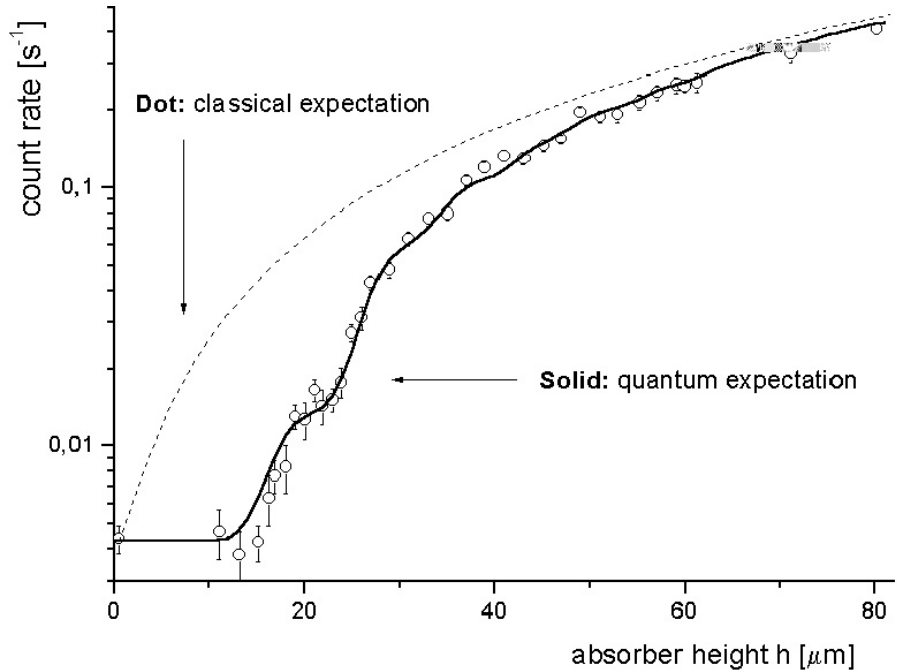}
\end{center}
\caption[]{Data and quantum expectation} \label{eps2}
\end{figure}
At sufficiently high slit width one approaches the classical
dependence and the stepwise increase is washed out. Fig. 6 shows
details of the quantum regime below an absorber height of  $h$ =
50 $\micro$m. The data follow this expectation as described: No
neutrons reach the detector below an absorber height of
15$\micro$m as explained before. Then above an absorber height of
15$\micro$m, we expect the transmission of ground-state neutrons
resulting in an increase in count rate. The expectation in this
case is shown in a solid line and agrees nicely with the data. The
$\chi^2$ is 56 for 35 degrees of freedom for one fit parameter,
the neutron flux~\cite{Westphal1}. The expectation for neutrons
behaving as classical particles is shown in a dotted line. The
classical expectation for neutron transmission is in clear
disagreement with the data (open circle). Especially, no neutrons
are transmitted for an absorber height between zero and fifteen
micrometer.
\section{Summary}
In this experiment, gravitational bound quantum states have been
seen for the first time. The experiment shows that, under certain
conditions, neutrons do not follow the classical Galileian
expectation when reflected from neutron mirrors. The measurement
does well agree with a simple quantum mechanical description of
quantum states in the earth's gravitational field together with a
mirror-absorber system. We conclude that the measurement is in
agreement with a population of quantum mechanical modes. Further,
the spectrometer operates on an energy scale of pico-eVs and
suitably prepared mirrors can usefully be employed in measurements
of fundamental constants or in a search for non-Newtonian gravity.
The present data constrain Yukawa-like effects in the range
between 1 $\micro$m and 10 $\micro$m. The limit for strength
$\alpha$ at 10 $\micro$m is 10$^{11}$ and at 1 $\micro$m the limit
is 10$^{12}$. This work has been funded in part by the German
Federal Ministry (BMBF) under contract number 06 HD 854 I and by
INTAS under contract number 99-705.

\end{document}